# Scattering of singular beams by subwavelength objects

Evyatar Hemo,* Boris Spektor, and Joseph Shamir

Department of Electrical Engineering, Technion—Israel Institute of Technology, Haifa 32000, Israel

*Corresponding author: sevyatar@012.net.il

In recent years, there has been a mounting interest in better methods of measuring nanoscale objects, especially in fields such as nanotechnology, biomedicine, cleantech, and microelectronics. Conventional methods have proved insufficient, due to the classical diffraction limit or slow and complicated measuring procedures. The purpose of this paper is to explore the special characteristics of singular beams with respect to the investigation of subwavelength objects. Singular beams are light beams that contain one or more singularities in their physical parameters, such as phase or polarization. We focus on the three-dimensional interaction between electromagnetic waves and subwavelength objects to extract information about the object from the scattered light patterns.   © 2010 Optical Society of America

## 1. Introduction

Recent progress in science and technology has increased the demand for reliable analysis of nanoscale structures. Measuring methods to sense and analyze smaller and smaller objects have been constantly pushed to their limits. While many high-quality laboratory methods exist, industrial applications also require high speed and ease of use, which is not common for most existing methods. Most practiced approaches for nanoscale inspection are based on scanning (confocal) microscopy, scanning probe microscopy, scanning electron microscopy, and interferometry. Each of these approaches has its intrinsic advantages and disadvantages in terms of resolution, working distance, operation speed, the need for special sample preparation, system complexity, and convenience of use. For modern technological applications, it is important to have high resolution as well as high speed, which are not compatible in most current methods. This limitation can be mitigated with the approach of singular beam microscopy [1], where a singular beam is used for illumination in a scanning microscopic system. Singular beams are beams that possess one or more singularities in their parameters, such as phase or polarization [2,3]. Typical examples of singular beams are the Gauss–Laguerre (GL) beams of different orders. While the zero-order GL beam is the fundamental Gaussian beam, higher order GL beams have a phase singularity at their center. Because the phase is not defined at the center, the amplitude must vanish, generating a doughnut-shape intensity distribution. This "dark hole" is obviously narrower than the beam width, and this property can be exploited for high-resolution inspection. Another type of singular beam is the dark beam (DB), which is a beam possessing a $\pi$ phase jump along an arbitrary line.

Earlier experimental studies using DB illumination for the detection and measurement of particles, defects, and simple structures on surfaces indicated extremely high resolution [1,4–6]. In this paper, we restrict the discussion for spheres embedded in homogeneous media and leave the analysis of the more complicated case of particles on surfaces [7] to a later work. The limitation to particles in homogeneous media still encompasses a large variety of applications, such as monitoring atmospheric pollution [8], pharmacology, biology, water technology, and the paint industry.

The main objective of this work is to gain an understanding of the basic interaction between singular beams and subwavelength objects, while keeping in mind the practical aspects of system design for the analysis of particle distributions. Unlike most conventional particle analysis instruments that are based on light scattering by particle ensembles, the optical system addressed here is designed for single particle measurements [9].

The conventional approach to treat the interaction between light and small particles is the Mie scattering theory [10,11]. Because that theory is valid only for plane wave illumination, we have to develop a generalized version suitable for arbitrary illuminating beam structures. As we are interested in the nanoscale domain, we must also deal with high numerical aperture (NA) optics, and, therefore, we present in the next section a theoretical background of vector scattering theory with arbitrary illumination. This is closely related to a previous work [12], which had already shown interesting results for scattering simulations, but it was devoted to GL beams only. Moreover, this work also addresses the design of the detection system and considers the effects of noise that may originate from various sources.

Section 3 discusses the representation of singular beams, and Section 4 presents a numerical study of DB scattering. In Section 5 we provide a comparison between scattering of Gaussian, GL, and DBs to demonstrate the superior performance of the DB in the current application. Concluding remarks are provided at the end.

## 2. Theoretical Background—Generalized Scattering Theory

### A. Scattered Electromagnetic Field Components

The classical approach to study the scattering of electromagnetic waves by small spheres is based on the Mie theory [10,11]. However, because the Mie theory was developed for plane wave illumination, it must be generalized for illumination with more complex beam structures, such as singular beams. We shall represent the scattered field as a superposition of TM and TE waves defined with respect to the radial direction, such that $H_r = 0$ and $E_r = 0$, respectively. Following the Bromwich formulation [10,13], the electric and magnetic fields are derived, respectively, from the two Bromwich potentials, also known as the Hertz–Debye potentials [14]. There are two potentials, $U_{TM}$ and $U_{TE}$, each of which must fulfill the wave equation. Using the spherical coordinate system, the wave equation has the form

$$\frac{\partial^2 U}{\partial r^2} + k^2 U + \frac{1}{r^2 \sin(\theta)} \frac{\partial}{\partial \theta}\left(\sin(\theta) \frac{\partial U}{\partial \theta}\right) + \frac{1}{r^2 \sin^2(\theta)} \frac{\partial^2 U}{\partial \varphi^2} = 0, \quad (1)$$

As indicated in Fig. 1, we assume a spherical scatterer, and, for the Bromwich formalism, the origin of the coordinate system $O$ is chosen at its center on the $x$–$y$ plane. Starting with the beam center also at $O$, it can be displaced to simulate relative object motion.

The solutions of Eq. (1) for the two potentials, $U_{TM}$ and $U_{TE}$, are given by

$$U_{TM} = \frac{E_0}{k} \sum_{n=1}^{\infty} \sum_{m=-n}^{+n} c_n g_{n,TM}^m \\ \times \exp(im\varphi) \begin{Bmatrix} \Psi_n(kr) \\ \zeta_n(kr) \end{Bmatrix} P_n^{|m|}(\cos\theta), \quad (2)$$

$$U_{TE} = \frac{H_0}{k} \sum_{n=1}^{\infty} \sum_{m=-n}^{+n} c_n g_{n,TE}^m \\ \times \exp(im\varphi) \begin{Bmatrix} \Psi_n(kr) \\ \zeta_n(kr) \end{Bmatrix} P_n^{|m|}(\cos\theta), \quad (3)$$

where $k$ is the wavenumber, $k = M\frac{\omega}{c}$, $\omega$ is the angular frequency, $c$ is the speed of light, and $M$ is the sphere complex refractive index. The Bromwich beam coefficients of a plane wave, $c_n$, are given by

$$c_n = \frac{1}{ik}(-i)^n \frac{2n+1}{n(n+1)}. \quad (4)$$

and the structure of a beam is described by the beam shape coefficients, $g_{n,TM}^m$ and $g_{n,TE}^m$, that are 1 for a plane wave. The functions $\Psi_n(kr)$ and $\zeta_n(kr)$ are the Ricatti–Bessel functions, defined by

$$\Psi_n(kr) = kr \cdot j_n(kr) = \left(\frac{\pi kr}{2}\right)^{\frac{1}{2}} J_{n+\frac{1}{2}}(kr), \quad (5)$$

$$\zeta_n(kr) = kr \cdot h_n^{(2)}(kr) = \left(\frac{\pi kr}{2}\right)^{\frac{1}{2}} H_{n+\frac{1}{2}}^{(2)}(kr), \quad (6)$$

where $H_n^{(2)}(kr)$ is a superposition of the Bessel and Neumann functions and is called a Hankel function of the second kind. The Hankel function has an important property of vanishing when $kr \to \infty$. $P_n^{|m|}(\cos\theta)$ are the well-known associated Legendre polynomials.

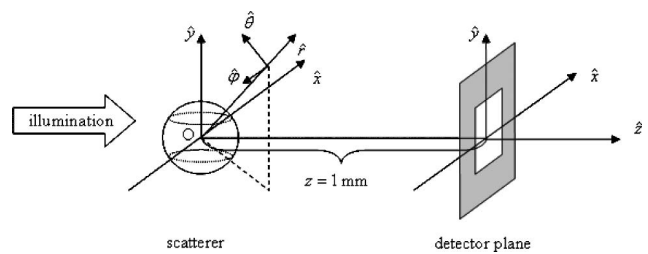

Fig. 1. Definition of the coordinate system.

After $U_{\text{TM}}$ and $U_{\text{TE}}$ are determined, the field components can be calculated by the following expressions [11]:

$$E_{r,\text{TE}} = 0, \qquad E_{\theta,\text{TE}} = -\frac{i\omega\mu}{r\sin\theta}\frac{\partial U_{\text{TE}}}{\partial\varphi},$$
$$E_{\varphi,\text{TE}} = \frac{i\omega\mu}{r}\frac{\partial U_{\text{TE}}}{\partial\theta}, \qquad (7)$$

$$E_{r,\text{TM}} = \frac{\partial^2 U_{\text{TM}}}{\partial r^2} + k^2 U_{\text{TM}}, \quad E_{\theta,\text{TM}} = \frac{1}{r}\frac{\partial^2 U_{\text{TM}}}{\partial r\partial\theta},$$
$$E_{\varphi,\text{TM}} = \frac{1}{r\sin\theta}\frac{\partial^2 U_{\text{TM}}}{\partial r\partial\varphi}, \qquad (8)$$

$$H_{r,\text{TM}} = 0, \qquad H_{\theta,\text{TM}} = \frac{i\omega\epsilon}{r\sin\theta}\frac{\partial U_{\text{TM}}}{\partial\varphi},$$
$$H_{\varphi,\text{TM}} = -\frac{i\omega\epsilon}{r}\frac{\partial U_{\text{TM}}}{\partial\theta}, \qquad (9)$$

$$H_{r,\text{TE}} = \frac{\partial^2 U_{\text{TE}}}{\partial r^2} + k^2 U_{\text{TE}}, \quad H_{\theta,\text{TE}} = \frac{1}{r}\frac{\partial^2 U_{\text{TE}}}{\partial r\partial\theta},$$
$$H_{\varphi,\text{TE}} = \frac{1}{r\sin\theta}\frac{\partial^2 U_{\text{TE}}}{\partial r\partial\varphi}. \qquad (10)$$

Using these relations, it is possible to derive the total field components as

$$E_r = -kE_0 \sum_{n=1}^{\infty}\sum_{m=-n}^{+n} c_n g_{n,\text{TM}}^m a_n[\zeta_n''(kr) + \zeta_n(kr)]P_n^{|m|}(\cos\theta)\exp(im\varphi), \qquad (11)$$

$$E_\theta = -\frac{E_0}{r}\sum_{n=1}^{\infty}\sum_{m=-n}^{+n} c_n[g_{n,\text{TM}}^m a_n \zeta_n'(kr)\tau_n^{|m|}(\cos\theta) + mg_{n,\text{TE}}^m b_n\zeta_n(kr)\Pi_n^{|m|}(\cos\theta)]\exp(im\varphi), \quad (12)$$

$$E_\varphi = -\frac{iE_0}{r}\sum_{n=1}^{\infty}\sum_{m=-n}^{+n} c_n[mg_{n,\text{TM}}^m a_n \zeta_n'(kr)\Pi_n^{|m|}(\cos\theta) + g_{n,\text{TE}}^m b_n\zeta_n(kr)\tau_n^{|m|}(\cos\theta)]\exp(im\varphi), \quad (13)$$

where $\tau_n^m$ and $\Pi_n^m$ are defined by

$$\tau_n^m(\cos\theta) = \frac{d}{d\theta}P_n^m(\cos\theta), \qquad (14)$$

$$\Pi_n^m(\cos\theta) = \frac{P_n^m(\cos\theta)}{\sin\theta}. \qquad (15)$$

The scattering process is taken into account by the Mie coefficients $a_n$ and $b_n$, [11], which are expressed by

$$a_n = \frac{\Psi_n(x)\Psi_n'(y) - M\Psi_n'(x)\Psi_n(y)}{\zeta_n(x)\Psi_n'(y) - M\zeta_n'(x)\Psi_n(y)}, \qquad (16)$$

$$b_n = \frac{M\Psi_n(x)\Psi_n'(y) - \Psi_n'(x)\Psi_n(y)}{M\zeta_n(x)\Psi_n'(y) - \zeta_n'(x)\Psi_n(y)}, \qquad (17)$$

where $x = ka = \frac{2\pi a}{\lambda}, y = k_{\text{sp}}a$, with $a$ being the sphere radius and $k_{\text{sp}}$ the wavenumber inside the sphere (may be complex for absorbing spheres). The refractive index of the sphere is given by $M = \frac{k_{\text{sp}}}{k}$. The function $\Psi_n(kr)$ is one of the Ricatti–Bessel functions corresponding to a first-order Bessel function.

It can be seen from Eqs. (11)–(13), that, for a given incident illumination,

$$E_r^{\text{incident}} = \frac{E_0}{kr^2}\sum_{n=1}^{\infty}\sum_{m=-n}^{+n} c_n g_{n,\text{TM}}^m n(n+1)\Psi_n(kr)$$
$$\times P_n^{|m|}(\cos\theta)\exp(im\varphi), \qquad (18)$$

$$H_r^{\text{incident}} = \frac{H_0}{kr^2}\sum_{n=1}^{\infty}\sum_{m=-n}^{+n} c_n g_{n,\text{TE}}^m n(n+1)\Psi_n(kr)$$
$$\times P_n^{|m|}(\cos\theta)\exp(im\varphi), \qquad (19)$$

$a_n$ and $b_n$ are the only differences between the incident and the scattered fields. Hence, the scattered field is completely determined by the Mie coefficients that depend on the sphere radius and refractive index, so that the difference between the scattered fields of two different materials can be represented by these coefficients.

### B. Shape of the Incident Beam

As indicated above, the beam shape is uniquely represented by the beam shape coefficients, $g_{n,\text{TM}}^m$ and $g_{n,\text{TE}}^m$. With the help of the orthogonality properties of the Legendre polynomials and complex exponential functions, it is possible to calculate the beam shape coefficients using the electric and magnetic radial fields. Applying the orthogonality properties

$$\int_0^{2\pi} \exp[i(m-m')\varphi]d\varphi = 2\pi\delta_{mm'}, \qquad (20)$$

$$\int_0^{\pi} P_n^m(\cos\theta)P_{n'}^m(\cos\theta)\sin\theta d\theta = \frac{2}{2n+1}\frac{(n+m)!}{(n-m)!}\delta_{nn'}, \qquad (21)$$

to Eq. (18) yields

$$\int_0^\pi P_n^{|m|}(\cos\theta)\sin\theta d\theta \int_0^{2\pi} \exp(-im\varphi)\frac{E_r^{\text{incident}}}{E_0}d\varphi$$
$$= \frac{1}{i(kr^2)}(-i)^n(2n+1)\Psi_n(kr)g_{n,\text{TM}}^m \frac{4\pi}{2n+1}\frac{(n+m)!}{(n-m)!}. \quad (22)$$

After some algebra, we obtain

$$g_{n,\text{TM}}^m = \frac{i^{n+1}}{4\pi}\frac{kr}{j_n(kr)}\frac{(n-m)!}{(n+m)!}\int_0^\pi P_n^{|m|}(\cos\theta)\sin\theta d\theta$$
$$\times \int_0^{2\pi}\exp(-im\varphi)\frac{E_r^{\text{incident}}}{E_0}d\varphi. \quad (23)$$

In a similar way, we can apply the orthogonality properties to the magnetic field in Eq. (18) and obtain an expression for the $g_{n,\text{TE}}^m$:

$$g_{n,\text{TM}}^m = \frac{i^{n+1}}{4\pi}\frac{kr}{j_n(kr)}\frac{(n-m)!}{(n+m)!}\int_0^\pi P_n^{|m|}(\cos\theta)\sin\theta d\theta$$
$$\times \int_0^{2\pi}\exp(-im\varphi)\frac{H_r^{\text{incident}}}{H_0}d\varphi. \quad (24)$$

## 3. Singular Beams

As indicated above, we refer to light beams that contain one or more singularities as singular light beams [10,11]. Such singularities are usually points or lines where one or more physical parameters, such as phase or polarization, are undefined. In this work we study two different types of singular beams: the GL beam and the DB.

### A. Gauss–Laguerre Beams

The higher than zero-order GL modes possess a helical phase that generates a phase singularity at their center. This phase singularity produces a narrow dark region at the center of the corresponding GL mode, which can be narrower than the light beam at the diffraction limit. The main objective of this work is to show that this property can be exploited to retrieve information regarding small nanoscale objects. Neglecting the time dependence, the GL modes [15,16] can be represented by

$$u_{nm}(r) = G(\tilde{\rho},\tilde{z})R_{nm}(\tilde{\rho})\Phi_m(\phi)Z_n(\tilde{z}), \quad (25)$$

where $\tilde{\rho} = \frac{\rho}{\omega(\tilde{z})}$ is the radial coordinate scaled by the spot size, $\omega(\tilde{z}) = \omega_0[1+\tilde{z}^2]^{1/2}$, $\tilde{z} = \frac{z}{z_0}$ is the longitudinal coordinate scaled by the Rayleigh length, $z_0 = \frac{\pi\omega_0^2}{\lambda}$, and

$$G(\tilde{\rho},\tilde{z}) = \frac{\omega_0}{\omega(\tilde{z})}\exp(-\tilde{\rho}^2)\exp(i\tilde{\rho}^2\tilde{z})\exp[-i\psi(\tilde{z})], \quad (26)$$

$$R_{nm}(\tilde{\rho}) = \left(\sqrt{2}\tilde{\rho}\right)^{|m|}L_{(n-|m|)/2}^{|m|}(2\tilde{\rho}^2), \quad (27)$$

$$\Phi_m(\phi) = \exp(im\phi), \quad (28)$$

$$Z_n(\tilde{z}) = \exp[-in\psi(\tilde{z})], \quad (29)$$

where $\psi(\tilde{z}) = \arctan(\tilde{z})$ and $L_{(n-|m|)/2}^{|m|}$ are the generalized Laguerre polynomials, where the integers $n$ and $m$ satisfy

$$n = |m|, |m|+2, |m|+4, |m|+6\ldots \quad (30)$$

It is noteworthy that the function $G(\tilde{\rho},\tilde{z})$ represents the radial Gaussian envelope of the beam and is independent of $n$ and $m$. A different commonly used notation for a GL beam is $G_{\frac{n-|m|}{2},m} = u_{nm}$. The helical phase and its singularity can be observed in Fig. 2.

### B. Dark Beam

A different type of singular beam, which has a $\pi$ phase jump along a single line, is the DB. The beam is called "dark" due to a narrow dark line along its center, which can be observed in Fig. 3. One possible way to generate such a beam is by a superposition of two GL beams with opposite modes, as given by

$$D = \frac{GL_{0,1} - GL_{0,-1}}{2}. \quad (31)$$

Considering the GL beam representation shown in Eq. (25), it is obvious that, in the corresponding expression for $D$, the factors $G$, $R_{nm}$ and $Z_n$ remain the same as for the GL beam discussed in the earlier sections. The only factor to be changed is the $\Phi_m$ factor, which, instead of creating the GL beam helical phase, turns into $\cos(\phi)$ [or $\sin(\phi)$ in case we add the two modes]. The consequence of the $\cos(\phi)$ factor is a beam with an opposite phase on its two sides, creating a narrow dark region along its center. As indicated above, the total width of the DB is determined by the Gaussian envelope having the width of $\omega_0$, while the distance between the two maxima of the DB is $\omega_0 \sqrt{2}$.

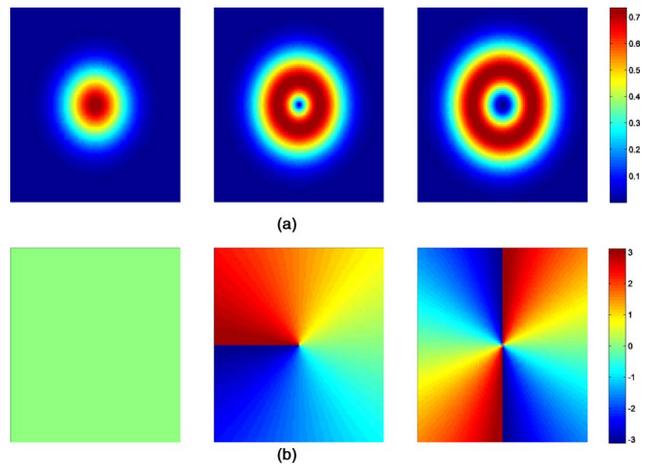

Fig. 2. (Color online) Field component distribution, $E_x$ of an $x$-polarized incident beam: (a) amplitude and (b) phase for three GL beams—amplitude scale is arbitrary and phase scale is in radians (white or red represents high intensity and black or blue represents low intensity).

## 4. Dark Beam Scattering—Numerical Study

### A. Configuration of Detectors

To assess the ability of nanosphere detection, the following case study was performed. A sphere of radius $a = 100 \, \text{nm}$ was illuminated by a DB with $\lambda = 405 \, \text{nm}$, $\omega_0 = 1 \, \mu\text{m}$ (NA = 0.125). Four different material compositions of the sphere were investigated: glass, GaAs, gold, and aluminum with the respective complex refractive indices of 1.5, $4.434 + 2.052i$, $1.658 + 1.956i$, and $0.503 + 4.923i$ [17]. The scattered field was calculated 1 mm from the scatterer at what will be defined as the detection plane (DP). To model a sphere moving in the transverse direction, the simulated illuminating beam center was positioned at different distances from the center of the sphere along the $x$ axis. The displacement interval was $0.25 \, \mu\text{m}$. To assess the detection capabilities, several detector configurations were examined. Keeping in mind the practicality of a system implementation, we consider here only configurations for forward detection, but the results provide some implications regarding other possibilities as well. For each configuration, the size and structure of the detectors were optimized so that the detector output will show maximal change between the presence and the absence of the scatterer.

### 1. Detector at Beam Center

Because of the dark center of the beam and the small size of the sphere relative to the beam width, a nanosphere situated exactly at the center will scatter an insignificant fraction of the illuminating beam. As a consequence, a $10 \, \mu\text{m} \times 10 \, \mu\text{m}$ detector placed at the center of the beam, also in the dark region over the DP, will measure a negligibly small integrated intensity. However, as the scatterer moves away from the center, it interacts with the illuminating beam, scattering light into the detector. The scattered light is now detected by the detector, as would be in an on-axis dark-field microscope. In fact, because we deal with small particles, the Mie scattering is close to isotropic and, therefore, the signal in this detector will closely resemble that obtained from a similar detector positioned at 90° or 80° to the optical propagation axis, which is technically more complicated.

The values of the incident field were normalized in such a way that the entire energy of the incident beam was 1. As a consequence, the maximal value of power of the incident beam at a single pixel (size of $2.5 \, \mu\text{m} \times 2.5 \, \mu\text{m}$) was $1.8 \times 10^{-4}$. The results shown in Fig. 4 indicate a change of $\sim 0.5 \times 10^{-6}$ between the minimal and maximal outputs of the detector. While this signal appears quite small, it still amounts to about 7% of the total detector signal. Unfortunately, as we shall see later, the presence of noise will dramatically affect this measurement.

### 2. Differential Measurement by Two Detectors at Both Sides of the Beam

One of the properties of the DB is its symmetric intensity distribution relative to the $y$ axis. This symmetry is broken whenever a scatterer displaced from the symmetry axis is present. This symmetry breaking can be measured by subtracting the integrated intensity over two detectors located symmetrically relative to the $y$ axis. In the simulation presented here, two $75\,\mu m \times 500\,\mu m$ detectors were positioned on both sides of the $y$ axis on the DP with a separation of $125\,\mu m$ between their centers. The distance between the detectors was determined so that their differential signal will have maximal change between the cases of the presence or absence of a scatterer. As expected, the specified distance put the detector centers at the points where the intensity gradient is maximal.

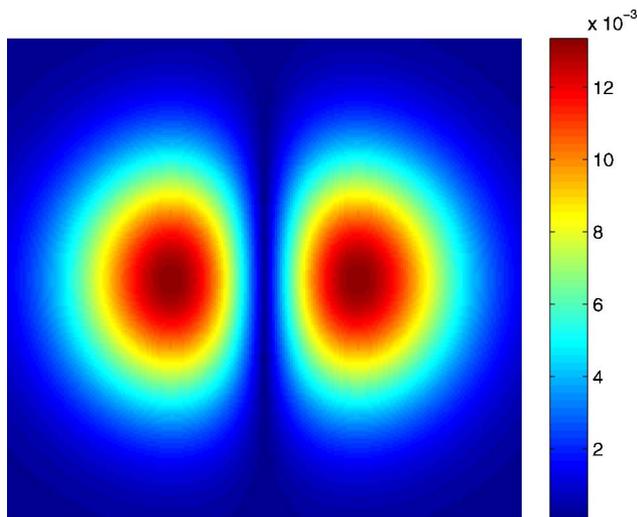

Fig. 3. (Color online) Field amplitude distribution of a DB; the amplitude scale is arbitrary. The two sides possess opposite phase.

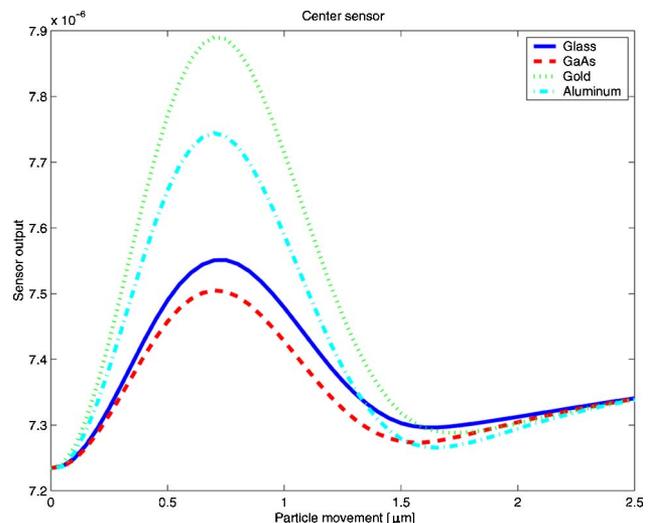

Fig. 4. (Color online) Center sensor output for moving sphere, $a = 100 \, \text{nm}$, NA = 0.125.

As Fig. 5 indicates, this detector configuration generates a much larger signal and, for a sphere crossing the whole beam, a similar but positive peak is obtained on the other side of the center. Because of its differential operation, this configuration is quite robust against noise, as will be shown below. The output of the detectors is again relative to the energy of the incident beam. It is worth noting that one of the reasons for the large signal is the fact that we detect the coherent superposition of the scattered signal with the direct beam that has an opposite phase on the two detectors. Moreover, in this configuration, much larger detectors can be utilized than for the central dark-field detector. For the present simulations, in the absence of a scatterer each detector integrated about 25% of the total power.

*3. Total Forward Integrated Intensity*

A $500\,\mu m \times 500\,\mu m$ detector measuring the total integrated intensity at the DP gave the results of Fig. 6. Because this detector practically integrates the whole power reaching the DP, its output is normalized so that the total integrated intensity with no scatterer is 1. The main contribution to this signal is the obstruction of the beam by the scatterer, and the calculated sensor output indicates a change of about 1%–3.5% of the total measured power. It is interesting to observe the significant difference between a pure dielectric (glass) and a conductor of the same size (gold and aluminum).

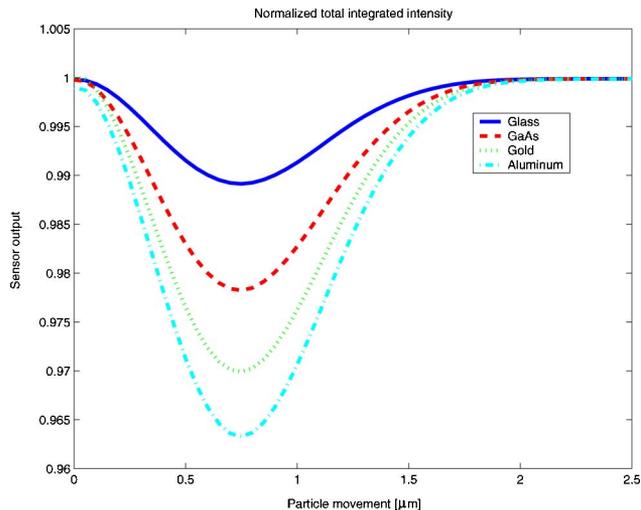

Fig. 6. (Color online) Total forward intensity-integrating detector output, $a = 100\,nm$, NA = 0.125.

### B. Higher NA Illumination

The above results were obtained with a relatively low NA (0.125) suitable for analysis within the paraxial approximation. While the above results were obtained with a full vector calculation, paraxial scalar methods would lead to similar results. The obvious aspect of higher NA beam illumination is a narrower focal spot with a still narrower dark region. This is also associated with a wider distribution over the DP. A less obvious aspect is the modification of the illuminating field structure. The vector character of the beam begins to play an important role and the most striking deviation from the scalar approximation is the possible appearance of a strong longitudinal field component.

The influence of higher NA beams on the measurements described above was studied using a conservative NA = 0.25 having a beam width of $\omega_0 = 0.5\,\mu m$.

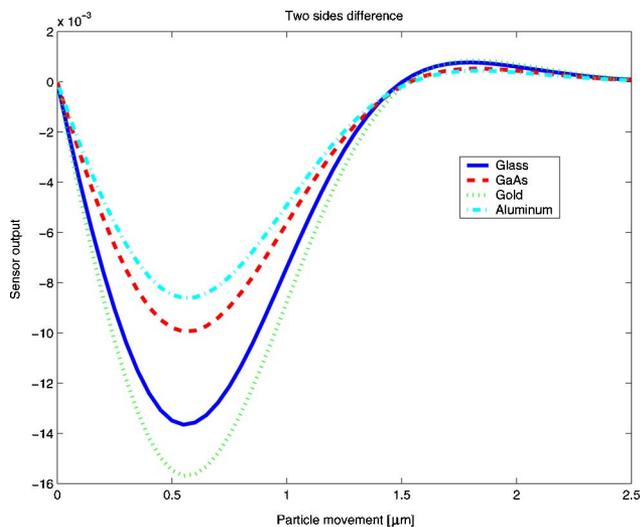

Fig. 5. (Color online) Differential detection with DB illumination, $a = 100\,nm$, NA = 0.125.

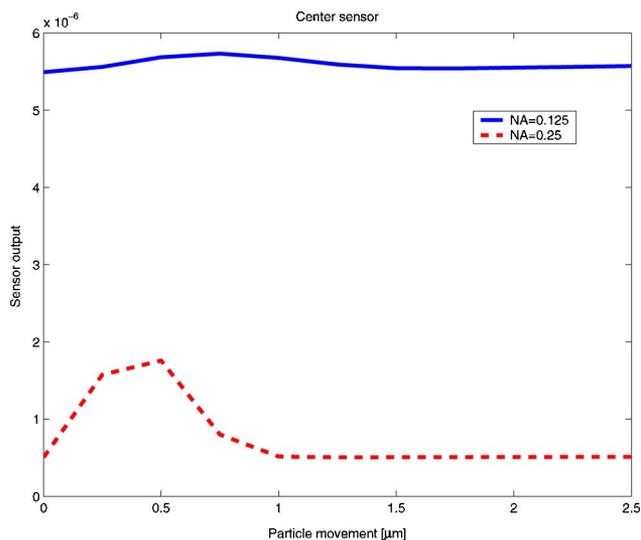

Fig. 7. (Color online) Comparison between the signal outputs from the central sensor for DB illumination with NA = 0.125 and NA = 0.25 ($a = 100\,nm$, $n = 1.5$).

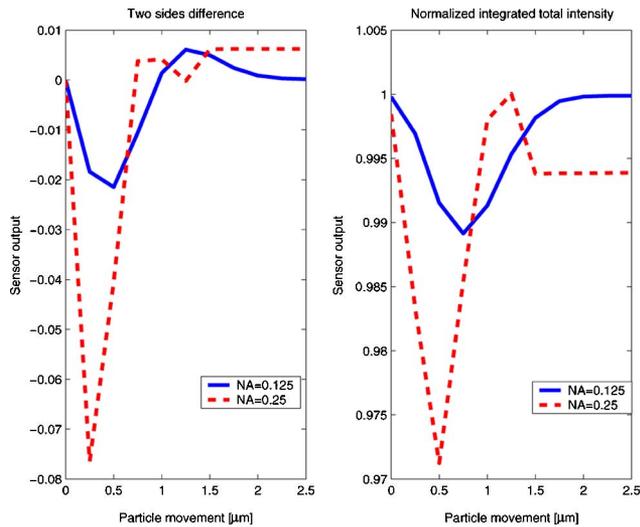

Fig. 8. (Color online) Same as Fig. 7 but for the differential detection configuration.

Similar to the previous section, the same three sensor configurations were examined with no alteration of the detector architectures. While some special effects due to the vector nature of the illumination are considered elsewhere [18], here we observe the three main effects resulting from the increased NA. First, for the central detector (Fig. 7) the background level of the higher NA measurement is much smaller. The reason for this effect is the wider dark region at the DP created by higher divergence of the narrower beam while the detector sizes were not changed. Second, the measurement changes as a result of the sphere presence for the higher NA scattering are more significant due to the fact that this beam is more concentrated. Third, the maximal measurement change is obtained for a smaller sphere displacement due to the fact that the incident beam is narrower.

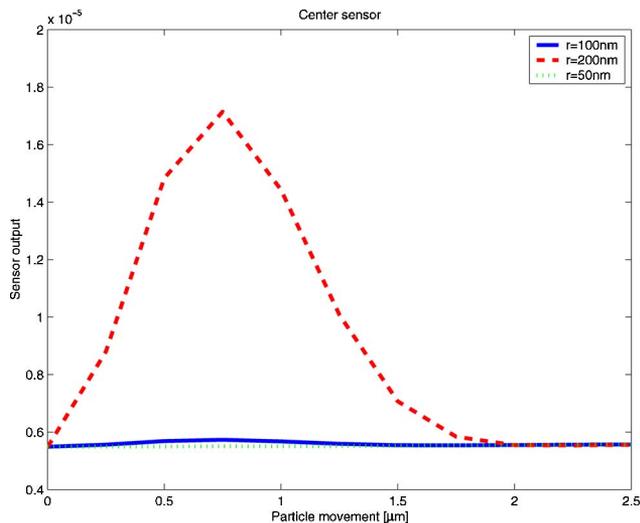

Fig. 9. (Color online) Comparison of center sensor output for DB illumination with NA = 0.125 for three sphere sizes ($a = 50, 100, 200$ nm, $n = 1.5$).

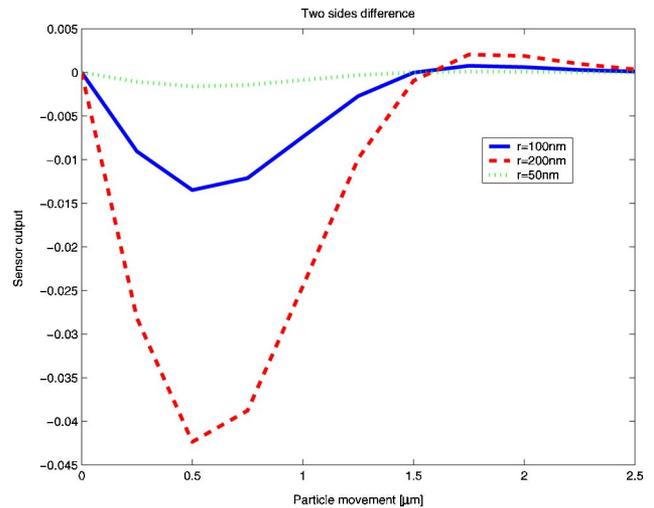

Fig. 10. (Color online) Same as Fig. 9 but for differential detection.

The sensor outputs for the rest of the sensor configurations are shown in Fig. 8. Similar to the center sensor configuration, the measurements are maximal for a smaller sphere displacement and the maximum value is much higher for the higher NA beam. It is again noteworthy that while the center and integral detector signals are symmetric around the beam center, the differential detector signal is antisymmetric, leading to double the signal shown for a full sphere crossing.

In principle, it is obvious that higher NA beams tend to improve the detection, but a practical implementation is more difficult.

### C. Scatterer Size Comparison

To assess the size discrimination capability, the system response was tested for different sphere sizes. Three spheres with radii of 50, 100, and 200 nm were tested using the same low-NA illumination as before (NA = 0.125) and the same three detector configurations.

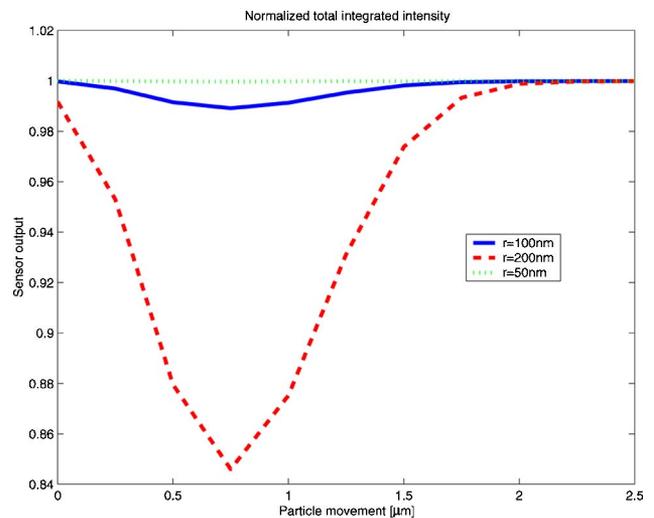

Fig. 11. (Color online) Same as Fig. 9 but for total power detection.

Measuring the light at the center of the detector plane gave a hardly noticeable signal for the 50 nm scatterer (Fig. 9). The signal obtained from the 200 nm sphere was ∼3 times larger than from the 100 nm sphere. As expected, in this region of scatterer sizes relative to beam size, the signal is monotonic with size but the prediction of the exact signal magnitude is not trivial. As can be seen from Figs. 10 and 11 the only detector configuration that could clearly sense the presence of the 50 nm sphere was the differential configuration.

## 5. System Response with Different Illuminating Beam Structures—Practical Aspects

In this section we address the practical aspects of the described system as an analytic tool. We start with a comparison of system response under a simple Gaussian beam illumination with the more complicated architecture using singular beam structures, the GL beam, and the DB. While the advantages of the DB are revealed for an ideal situation, these advantages are significantly enhanced for practical situations when noise is present in the system.

### A. Beam Comparison

To evaluate the system response for different illuminating beam structures, we simulated a conventional Gaussian beam, a GL beam, and a DB, all with exactly the same configuration and parameters as in Subsection 4.A. Unlike the previous figures, in Fig. 12 the complete beam crossing is shown. The smaller signal value for the GL beam can be explained by the fact that the energy of the GL beam is spread over a larger circular area and the intensity hitting the sphere is smaller than for the other two beams (Figs. 2 and 3). The DB shows a 70% improvement relative to the Gaussian beam in this measurement, which is a significant achievement for the inspection of nanoscale objects.

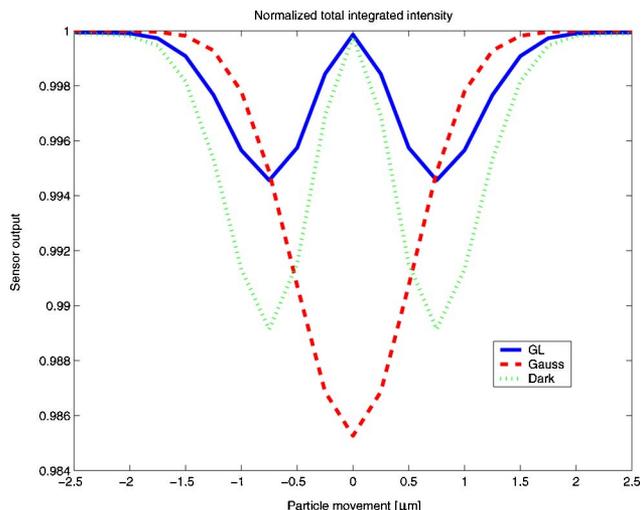

Fig. 13. (Color online) Total integrated intensity for the three beams, $a = 100$ nm, NA $= 0.125$, $n = 1.5$.

The total integrated intensity measurement output (Fig. 13) is roughly dependent on the fraction of absorbed energy and is monotonic with the incident beam concentration. The interaction of the most concentrated beam, which is the Gaussian beam, is the strongest, and, therefore, it yields the highest signal for this detector configuration. It should be noted at this point, however, that even the apparent large signal for a the Gaussian beam rides on a high bias level in contrast to the two other detector configurations that measure deviation from zero. There was no sense in comparing the center detector response because it will be overexposed by the Gaussian beam.

### B. Influence of Noise

The above results were obtained with the assumption of ideal circumstances. In practical applications, however, background noise is imminent. In order to examine the influence of noise on the detector measurement sensitivity, two kinds of noisy scenarios were simulated. One deals with static noise, such as the optical noise generated by scattering and reflections from static surfaces, and the second scenario models the scattering by moving objects such as particles not in the measurement volume and moving surface that may exist in the system.

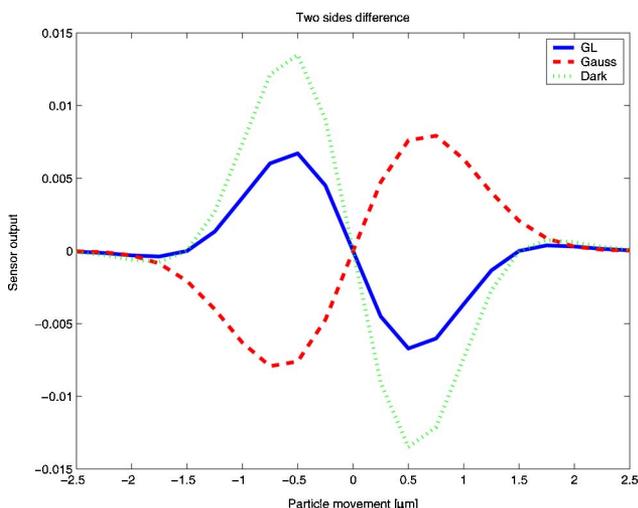

Fig. 12. (Color online) Comparison of the differential sensor for illumination with various beam structures, $a = 100$ nm, NA $= 0.125$, $n = 1.5$.

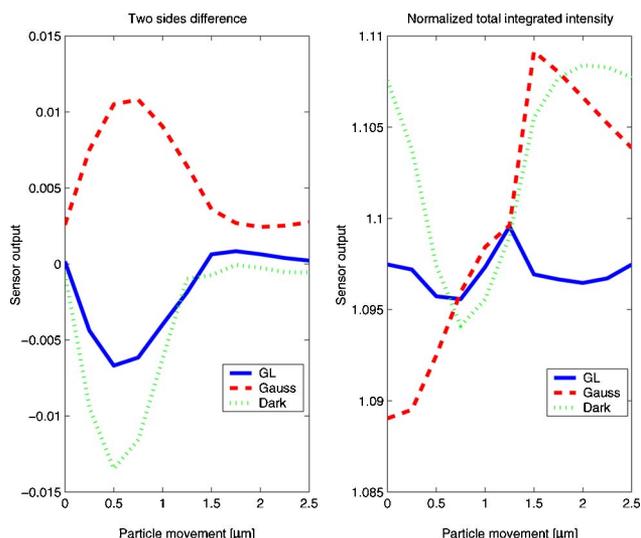

Fig. 14. (Color online) Signal comparison for the three illuminating beams in the presence of static noise, $a = 100$ nm, NA $= 0.125$, $n = 1.5$.

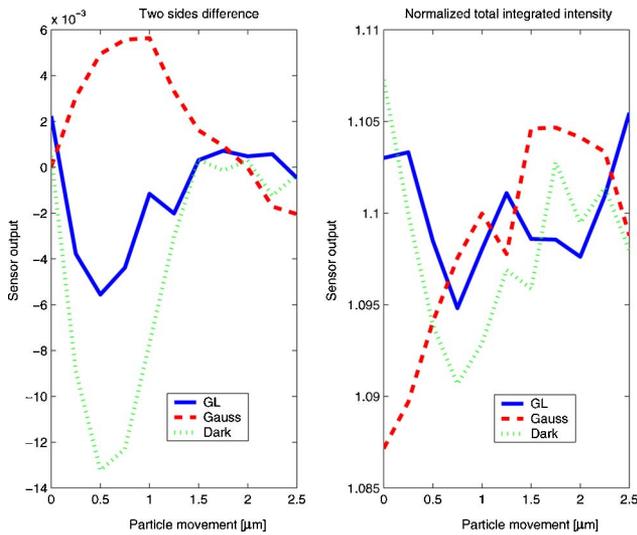

Fig. 15. (Color online) Same as Fig. 14 but with semidynamic noise.

*1. Static Noise*

Gaussian random noise was generated and coherently added to the total electric field at the detector plane. Such a noise is a good representation of light scattered off the optical components. We call this "static noise" because the same noise was added to all measurements. As an example, this noise was simulated with a variance such that the total energy of the noise at the DP was 10% of the total energy of the beam at this plane.

Obviously, the differential detector configuration is less sensitive to noise due to its differential operating mode. Thus, this configuration provides measurement results similar in their shape and signal-to-noise ratio (SNR) to the noiseless scenarios. The total intensity detector is more sensitive, and it is possible to see in Fig. 14 that the measurement bias and shape are changed due to the presence of the noise.

*2. Semidynamic Noise*

Static noise implies that the noise does not change during the measurements for different sphere positions. As indicated above, this only takes care of static scattering and some other constant background noise. In most practical situations, this noise is accompanied also by some dynamic noise, such as from the scattering of moving parts in the optical system and other objects that are not in the measurement region. A good model for this noise is a semidynamic noise where the correlation decreases with distance.

This is a fair model for a moving sphere when the noise changes with time. Let us denote the field of noise for sphere position $i$ by $n_i$, and then the correlation ratio for the noise is

$$\langle n_m, n_k \rangle = \left(\frac{1}{\sqrt{2}}\right)^{m-k}. \tag{32}$$

This means that the noise for two adjacent measurements is highly correlated, while for significantly separated measurements the noise is independent. The semidynamic noise simulation was also implemented with noise energy equivalent to 10% of the beam energy at the detector plane. The central sensor is very sensitive to this type of noise; its measurements are very noisy, and the presence of the sphere is completely obscured. The measurements for the other two sensor configurations can be seen in Fig. 15.

As expected, the measurements are noisy; however, a clear indication for the presence of the sphere can still be seen for the Gaussian and DBs. The differential detectors show even better improvement when the noise is not completely static, as was assumed in the previous simulations. The differential output of the DB shows a 130% improvement relative to the Gaussian one. This improvement can be explained by the fact that the interference caused by additive coherent noise is higher when it is added to higher values of the field, like in the Gaussian case.

It was interesting to test the measurement outputs for even smaller spheres. Therefore, semidynamic noise simulations were run for 50 nm spheres. Because, for these spheres, the 10% noise level was too high, in this simulation its total energy was reduced here to 2.5% of the beam energy at the DP, which is still a rather high noise level for practical situations. Figure 16 clearly demonstrates the advantages of the DB illumination as compared to the Gaussian beam illumination for the differential detection setup. In the measurements with the Gaussian beam, there is no indication for the presence of the sphere, whereas a clear signal is obtained with the DB configuration.

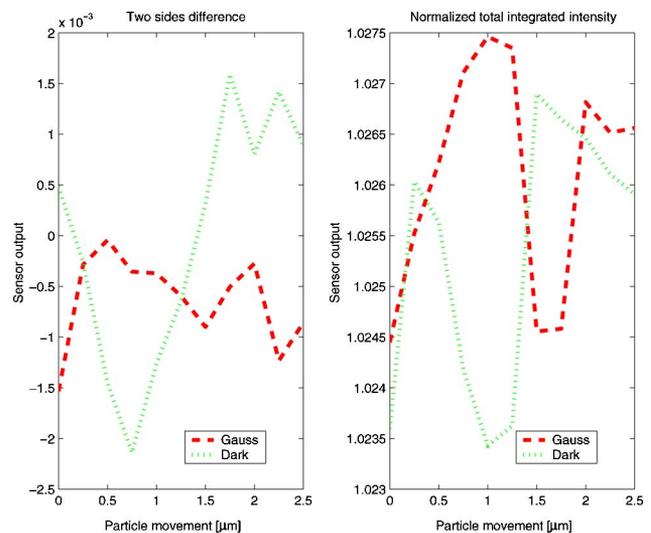

Fig. 16. (Color online) Same as Fig. 14 but with semidynamic noise for a 50 nm sphere.

*3. Shot Noise*

In addition to the optical noise considered above, it is of interest to also analyze the effect of the intrinsic detector noise. In general, shot noise in electronic devices consists of random fluctuations of the electric current in many electrical conductors, which are caused by the fact that the current is carried by electrons as well as the quantum detection process within the detectors. Shot noise is a Poisson process, and the charge carriers that make up the current will follow a Poisson distribution. The current fluctuations have a standard deviation of

$$\sigma_i = \sqrt{2eI\Delta F}, \qquad (33)$$

where $e$ is the electron charge, $\Delta F$ is the bandwidth of the detector, and $I$ is the average current.

As a consequence, when measuring intensity in the configurations described above, there is a difference whether the background of the desired signal is dark or not. For example, when measuring intensity by a small detector at the center of the beam, the indication for the presence of a sphere is surrounded by high-intensity measurements for the Gaussian beam case. However, for the GL and DBs, the background of the measured indicator is dark. Therefore, the SNR of the measurements for the GL and DB scenarios is expected to be higher in practical applications.

**6. Conclusions**

We have explored the scattering of singular beams by spheres in the nanoscale region while considering the practicality of a novel approach to analyze nanoparticle distributions in homogeneous media. The singular beams that were investigated included the GL beam and the DB with their respective point singularity (vortex) and line singularity. Several sensor configurations were offered in order to obtain information about the scatterer. The sensor configurations included a small sensor at the center of the beam, two adjacent differential sensors, and one large sensor. This paper demonstrated the sensitivity improvement gained by using DBs over using Gaussian beams. The differential detector shows the highest sensitivity improvement, which reaches a 70% SNR improvement. In order to simulate a more practical scattering scenario, we considered additional moving particles that scatter light but are not positioned within the measuring volume. The resulting optical noise was modeled by a coherent semidynamic noise added to the simulations. The semidynamic noise is a noise where the correlation decreases with the distance. The differential detector shows even higher SNR improvement using the DB when semidynamic noise is present. The predicted sensitivity for nanosphere inspection is 50 nm in the presence of quite significant noise. It is expected that this sensitivity can be improved further by additional system refinements.